\documentclass[12pt]{article}
\usepackage{graphicx}
\usepackage{dcolumn}
\usepackage{amsmath,amsthm,amssymb}
\usepackage{hyperref}
\usepackage{tikz}
\usetikzlibrary{calc}
\usetikzlibrary{arrows}
\usetikzlibrary{patterns}
\usepackage{float}
\begin{document}

\title{Three-dimensional gravity and instability of $\text{AdS}_{3}$}

\author{Joanna Ja\l mu\.zna \\
Faculty of Mathematics and Computer Science, \\
Jagiellonian University, \\
\L ojasiewicza 6, 30-348 Krak\'ow, Poland
}

\date{\today}
\begin{abstract}
This is an extended version of my lecture at the LIII Cracow School of Theoretical Physics in Zakopane in which I presented the results of joint work with Piotr Bizo\'n \cite{bj} concerning (in)stability of the three-dimensional anti-de Sitter spacetime.
\end{abstract}

\maketitle
%%%%%%%%%%%%%%%%%%%%%%%%%%%%%%%%%%%%%%%%%%%%%%%%%%%%%%%%%%%%
%%%%%%%  Introduction
%%%%%%%%%%%%%%%%%%%%%%%%%%%%%%%%%%%%%%%%%%%%%%%%%%%%%%%%%%%%

\section{Introduction}
General Relativity is the theory of space, time, and gravity formulated  by  Einstein in 1915. In this theory, the spacetime is a differential manifold  with a Lorentzian metric $g_{\alpha\beta}$ which satisfies the  Einstein equations
\begin{equation*} %\label{einstein}
R_{\alpha \beta}-\frac{1}{2} g_{\alpha \beta} R + \Lambda g_{\alpha \beta} = \kappa T_{\alpha \beta}.
\end{equation*}
 Here $R_{\alpha \beta}$ is a Ricci tensor, $R$ is a Ricci scalar, $T_{\alpha \beta}$ is the stress-energy tensor of matter, and $\Lambda$ is a cosmological constant. These equations form a very complicated  system of partial differential equations for the metric components.

The simplest, but physically very important, case one can consider is a vacuum solution describing an empty universe with no matter present ($T_{\alpha \beta}=0$). Among vacuum solutions there are three distinguished maximally symmetric spacetimes: Minkowski ($\Lambda=0$), de Sitter ($\Lambda>0$, and Anti-de Sitter (AdS) ($\Lambda<0$).
These solutions  play very important roles in theoretical physics. The Minkowski spacetime is a background for all physical theories in which gravitational effects can be neglected. The de Sitter solution describes an expanding universe and appears in cosmology both as a model of the early universe (inflation) and a candidate for the late time attractor. Finally, the AdS spacetime plays a prominent role  in the AdS/CFT correspondence which conjectures a duality between two physical theories: gravity  in the AdS bulk and  a conformal quantum field theory living on the AdS boundary.

An important question concerning these spacetimes is the question of their stability. The nonlinear stability was established for Minkowski by Christodoulou and Klainerman  in 1993 \cite{ck} and for de Sitter by Friedrich  in 1986 \cite{f}. However, for AdS the problem of stability  remains open. Some of the reasons why this problem is so difficult will be explained below.

%%%%%%%%%%%%%%%%%%%%%%%%%%%%%%%%%%%%%%%%%%%%%%%%%%%%%%%%%%%%
%%%%%%% Opis AdS, model, r�wnania, warunki brzegowe
%%%%%%%%%%%%%%%%%%%%%%%%%%%%%%%%%%%%%%%%%%%%%%%%%%%%%%%%%%%%

\section{Anti-de Sitter spacetime}
\begin{figure}
\centering
\begin{tikzpicture}

\draw[thick] (0,0) -- (0,5);
\draw[thick] (2,0) -- (2,5);

\draw [thick] (0,1) --  (2,3);
\draw [->,>=stealth,thick] (0.9,1.9) -- (1,2);

%\draw [dashed, thick] (2,3) -- (3,4);
\draw [->,>=stealth',dashed,thick] (2,3) -- (2.5,3.5);
\draw [dashed, thick] (2.5,3.5) -- (3,4);

%\draw [dashed,thick] (2,3) -- (1,4);
\draw [->,>=stealth',dashed,thick] (2,3) -- (1.5,3.5);
\draw [dashed,thick] (1.5,3.5) -- (1,4);

\draw[->,>=stealth']  (-0.5,2)--(-0.5,4);

\node [font=\Large,thick] at (2.3,3.0) {?};
\node [font=\large] at (-0.8,3) {t};
\node [font=\small] at (0,-0.5) {$x=0$};
\node [font=\small] at (2,-0.5) {$x=\frac{\pi}{2}$};
\end{tikzpicture}
\caption{A conformal diagram for AdS (the angular dimensions are suppressed). The line represents a light ray sent outwards from the center. The ray gets to infinity in time $t=\pi/2$. The question mark is meant to indicate that in order to determine the behaviour of the light ray (for example its polarization) for times $t>\pi/2$ one has to prescribe boundary conditions at  $\mathcal{I}$.}
\label{conf_diag}
\end{figure}
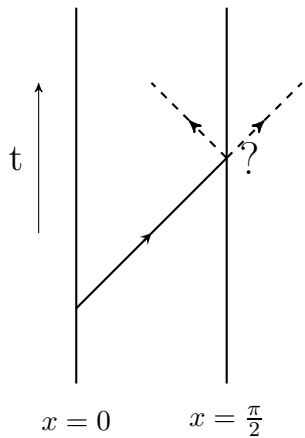

The $(d+1)$ dimensional Anti-de Sitter ($\textnormal{AdS}_{d+1}$) spacetime is the unique maximally symmetric solution to the vacuum Einstein equations with a negative cosmological constant. It is a Lorentzian analogue of the  hyperbolic space, just as Minkowski is an analogue of the Euclidean space. Geometrically, $\textnormal{AdS}_{d+1}$ can be defined as the $d+1$ dimensional quadric
$$
-U^2 -V^2 + \sum \limits_{i=1} ^{i=d} X_i^2 = -\ell^2
$$
embedded in the space with the metric
$$
ds^2=-dU^2-dV^2 + \sum \limits_{i=1} ^{d} dX_i^2.
$$
 Here $\ell$ is the length scale usually referred to as the AdS radius.
We introduce the parametrization
\begin{align*}
U&=\sqrt{r^2+ \ell^2} \sin (t/\ell), \\
V&=\sqrt{r^2+ \ell^2} \cos (t/\ell), \\
X &= r \omega,
\end{align*}
where $\omega$ is the parametrization of $d-1$ dimensional sphere. Then, the induced metric on the quadric takes the form
\begin{equation}\label{ads-metric}
  ds^2=-\left( 1+\frac{r^2}{\ell^2}\right) dt^2+\left( 1+\frac{r^2}{\ell^2}\right)^{-1} dr^2 +r^2 d \omega^2_{S^{d-1}},
\end{equation}
where $d \omega^2_{S^{d-1}}$ is the metric on the $(d-1)$-dimensional sphere. Simple calculation shows that the metric \eqref{ads-metric}  solves the vacuum Einstein equations with $\Lambda = -\frac{2}{d(d-1) \ell^2}$.  The time  coordinate $t$ is periodic but it can be 'unrolled' by passing to the universal covering space and then it has the range $t \in (-\infty, +\infty)$.  The AdS spacetime has some counterintuitive properties. To see that it is convenient to introduce a compactified variable $x=\arctan ( r/\ell) \in [0,\pi/2)$. Then the AdS metric \eqref{ads-metric} becomes
\begin{equation*} %\label{ads_x}
ds^2 = \frac{\ell^2}{\cos^2 x} \left( -dt^2 + dx^2 + \sin^2 x d\omega^2 _{S^{d-1}} \right).
\end{equation*}
From this one sees that AdS is conformal to one half (say, the northern hemisphere) of the Einstein static universe. The conformal boundary at  $x=\pi/2$, denoted hereafter by  $\mathcal{I}$, is the timelike cylinder $R\times S^2$. For better understanding of the causal structure of AdS, we show its conformal diagram  in Fig. \ref{conf_diag}.

The conformal infinity of AdS is timelike and null rays can reach it in finite proper time of the central observer. Therefore, in order to  determine the time evolution of waves propagating in this spacetime, it is not sufficient to specify initial data on a spacelike hypersurface $t=const$ but, in addition, one has to prescribe  boundary conditions at the conformal infinity $\mathcal{I}$.  In other words,  AdS is not globally hyperbolic.

 It follows from the above that the mechanism reponsible for the stability of the Minkowski spacetime - dissipation of energy by dispersion - is absent in AdS, because for no flux boundary conditions $\mathcal{I}$ acts as a mirror and energy cannot escape from it. The perturbations propagating outwards bounce back to the bulk leading  to complicated nonlinear wave interactions understanding of which is the key to the problem of stability of AdS.

 A first step towards understanding this problem was done by Bizo\'n and Rostworowski  \cite{br} who studied numerically and perturbatively  a self-gravitating massless scalar field in $3+1$ dimensional asymptotically AdS spacetime. They showed that for a  large class of arbitrarily small perturbations of $\text{AdS}_{4}$  a black hole forms after a time of order $\mathcal{O}(\varepsilon^{-2})$, where $\varepsilon$ is the size of initial data.  On the basis of nonlinear perturbation analysis they conjectured that this instability is due to the resonant transfer of energy from low to high frequencies.  This mechanism is reminiscent of  the turbulent energy cascade observed in fluids dynamics. Concentration of energy on finer spatial scales leads to  gravitational collapse (black hole formation is the analogue of the viscous cut-off for fluids). Later, this result was extended in a straightforward manner to  higher dimensions $d > 3$  \cite{jrb} and verified and generalized to more general models \cite{bll,bll2,mr,dhs}. In particular, the existence of time-periodic solutions was shown in \cite{mr,dhs}.

Somewhat surprisingly, the analogous problem in three dimensions seems more difficult. In the following I will present the results of my recent joint work with Piotr Bizo\'n \cite{bj} in which we studied numerically the evolution  of small perturbations $\text{AdS}_{3}$. Before presenting this work, I discuss some distinctive features of three-dimensional gravity which are necessary in understanding why the problem of stability of $\text{AdS}_{3}$ is different from the analogous problem in higher dimensions.

%%%%%%%%%%%%%%%%%%%%%%%%%%%%%%%%%%%%%%%%%%%%%%%%%%%%%%%%%%%%
%%%%%%% grawitacja w 2+1 bez sta?ej kosmologicznej i z nia, BTZ
%%%%%%%%%%%%%%%%%%%%%%%%%%%%%%%%%%%%%%%%%%%%%%%%%%%%%%%%%%%%
\section{Three-dimensional gravity} \label{3dim}

 Although the three-dimensional gravity is not physical, it serves as an interesting toy model  which helps us to understand which features of general relativity depend crucially on our world being four-dimensional. It is also an interesting  playground for studies of quantum gravity.

 Let me first explain what makes Einstein equations in three dimensions special from the PDE viewpoint. As any scale invariant system,
Einstein equations can be classified according to the scaling properties of energy $E\rightarrow \lambda^\alpha E$ under the dilation $x\rightarrow x/\lambda$. The degree $\alpha$ defines the criticality class of the equation. The equation is said to be  subcritical if $\alpha<0$, critical if $\alpha=0$, and supercritical if $\alpha>0$.  This classification is a basis of a heuristic  principle according to which solutions of subcritical evolution equations are globally regular in time, whereas solutions of supercritical equation develop singularities in finite time. In the case of Einstein equations the degree $\alpha=d-2$, hence the dimension $d=2$ is critical. Another way to see that is to look at the dimension of the Newton constant: $[G]=M^{-1} L^{d-2}$. Thus, only in $d=2$ we have a mass scale given by $1/G$.

  To see why the three-dimensional gravity is special geometrically, recall that in $n$ dimensions the Riemann curvature tensor has $\frac{n^2(n^2-1)}{12}$ independent components while the Ricci tensor has $n(n-1)$ independent components.  Thus, in three dimensions both the  Riemann and Ricci tensors have 6 independent components and, in fact, the Riemann tensor can be expressed by the Ricci tensor. Consequently, it follows from Einstein equations  that spacetime is flat where no matter is present. In other words, there are no gravitational waves in three dimensional gravity.

The pioneering  studies of three-dimensional gravity (without the cosmological constant) were done 50 years ago by Staruszkiewicz \cite{star} and presented by him at the III Cracow School of Theoretical Physics in Zakopane in 1963.

\begin{figure}[h]
\begin{center}
\begin{tikzpicture}
\draw [<->,domain=225:315] plot ({1.7*cos(\x)}, {1.7*sin(\x)});
\draw [dotted,domain=225:-45] plot ({sqrt(8)*cos(\x)}, {sqrt(8)*sin(\x)});
%\filldraw[domain=0:180,pattern=north east lines,pattern color=black!20] plot ({sqrt(8)*cos(\x)}, {sqrt(8)*sin(\x)});
\filldraw[draw=none,pattern=north east lines,pattern color=black!20, draw opacity=0] (0,0)--(2,-2) arc (-45:225:2.8284271)--cycle;
\coordinate [label=above:M] (M) at (0,0);
\coordinate [label=below:A] (A) at (-2,-2);
\coordinate [label=below:B] (B) at (2,-2);
\draw (A) -- (M);
\draw (M) -- (B);
\draw[thick,dashed,->,>=stealth'] (-2.2,1.11244) -- (-1.9,-0.00717968) ;
\draw[thick,dashed] (-1.9,-0.00717968) -- (-1.5,-1.5);
\draw [domain=45:105] plot ({1*cos(\x)-1.5}, {1*sin(\x)-1.5});
\node at (-1.3,-1) {\footnotesize{$\alpha$}};

\draw[thick,dashed,->,>=stealth'] (1.5,-1.5) -- (1.9,-0.00717968);
\draw[thick,dashed]  (1.9,-0.00717968) -- (2.2,1.11244);
\draw [domain=75:135] plot ({1*cos(\x)+1.5}, {1*sin(\x)-1.5});
\node at (1.3,-1) {\footnotesize{$\alpha$}};
\node at (0,-1.1) {\small{deficit angle}};
 \end{tikzpicture}
 \caption{A point particle is denoted by $M$. A wedge is cut out of the plane and the edges  $MA$  and $MB$ are glued together to get a cone. The angle deficit corresponds to a mass of the particle. This shows that even though the spacetime is flat outside of $M$, the presence of the point mass globally influences the geometry. The motion of a test particle in a such geometry is illustrated with the dashed line. The points where this line meets the edges are identified after gluing.}
 \label{fig_angle}
 \end{center}
 \end{figure}
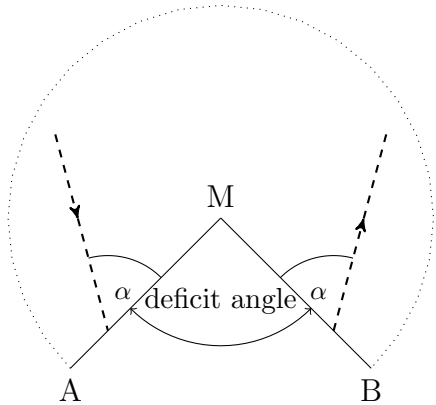

Staruszkiewicz solved the static $n$-body problem for a collection of point masses. His construction goes as follows.
A general static, circularly  symmetric metric has the form
$$ds^2=-e^{2A(r)} dt^2+e^{2B(r)} dr^2+r^2 d \varphi^2,$$
where $0 \leq r \leq +\infty$, $0 \leq \varphi < 2\pi$. A simple calculation shows that $R_{\mu \nu}=0$ implies that $A=const$, $B=const$. Without a loss of generality one can choose $A=0$. Then, the metric produced by a point particle is
\begin{equation} \label{metric_transf}
ds^2=-dt^2+e^{2B} dr^2+r^2 d \varphi^2,
\end{equation}
which by the transformation
$R=r e^B, \Phi=\varphi e^{-B}$ gives the flat metric
\begin{equation*}
ds^2=-dt^2+dR^2+R^2 d \Phi^2
\end{equation*}
 with the deficit angle: $0 \leq \Phi < 2 \pi e^{-B} < 2 \pi$ (assuming that $B > 0$, which corresponds to a positive mass). The spatial surface $t=const$  is a cone. It is obtained by cutting a wedge out of a Euclidean space and gluing the edges together. The missing part of the space is described by the deficit angle corresponding to the mass of the particle. Schematically this is shown in Fig. \ref{fig_angle}.

Staruszkiewicz showed that the above solution can be constructed geometrically. Let $ds^2=du^2+dv^2$ be the metric on the cone. In terms of the complex variable $w=u+i v$ this becomes $ds^2=dw d\bar{w}$. The transformation
$
w=z^{1-\frac{\alpha}{\pi}},
$
where $2 \alpha$ is the deficit angle connects the edges and brings the metric into the following form (in polar coordinates)
$$
ds^2=-dt^2+\left( 1-\frac{\alpha}{\pi} \right)^2 r^{-\frac{2 \alpha}{\pi}} (dr^2 +r^2 d\phi^2),
$$
where $0 \leq r \leq \infty$ and $0 \leq \phi < 2 \pi$. From this form of the metric it is obvious that
it is singular at $r=0$. By another coordinate transformation it can be shown that it is equivalent to the the metric \eqref{metric_transf}. This approach has the advantage that it can  be naturally generalized to the case of n point masses \cite{star} (see also \cite{thooft}).

% Subsection with preliminaries %%%%%%%%%%%%%%%%%%%%%%%%%%%
\subsection{Preliminaries}
 Following \cite{br,jrb}, in \cite{bj} we consider a simple asymptotically AdS solution in three dimensions, namely  a  self-gravitating spherically symmetric real massless scalar field. The dynamics of this solution is governed by the system of Einstein equations and the wave equation for the scalar field
 \begin{equation}\label{einstein-scalar}
G_{\alpha\beta}+\Lambda g_{\alpha\beta}= \kappa T_{\alpha\beta}\,,\quad  g^{\alpha\beta} \nabla_{\alpha}
\nabla_{\beta} \phi=0\,,
   \end{equation}
 where $T_{\alpha\beta}=
 \partial_{\alpha} \phi \partial_{\beta} \phi-\frac{1}{2} g_{\alpha\beta}(\partial \phi)^2$ is the stress-energy tensor of the scalar field and $\Lambda < 0$ is the cosmological constant. As a simplifying assumption, we assume the circular symmetry for the scalar field $\phi=\phi(\tau,r)$ and the metric
\begin{equation}\label{ansatz}
 ds^2 = -\tilde A e^{-2\delta}\, d\tau^2 + {\tilde A}^{-1} dr^2 + r^2\, d\varphi^2\,,
\end{equation}
where $\tilde A$ and $\delta$ are also functions of $(\tau,r)$ only. Substituting the ansatz \eqref{ansatz} into \eqref{einstein-scalar} we get
\begin{equation}\label{weq}
  \partial_{\tau} \left({\tilde A}^{-1} e^{\delta} \partial_{\tau} \phi\right)=\frac{1}{r}\partial_r  \left(r \tilde A e^{-\delta} \partial_r \phi\right)\,,
\end{equation}
for the wave equation and
\begin{eqnarray}
 \partial_r \delta &=& -\kappa\, r \,S\,, \label{slicing} \\
  \partial_r \tilde A  &=& -\kappa \,r \tilde A\, S - 2 \Lambda r, \label{hamilton0}\\
   \partial_{\tau} \tilde A &=& -2 \kappa\,  r \,\tilde A \,\partial_{
\tau} \phi\, \partial_r \phi\,, \label{momentum}
\end{eqnarray}
for the Einstein equations, where $S={\tilde A}^2 e^{-2\delta} (\partial_t \phi)^2+(\partial_r \phi)^2$. It is easy to check that in  a vacuum ($\phi=0$) this system has a family of static solutions
\begin{equation} \label{static_sol}
\delta=0, \quad \tilde A=1-M+r^2/\ell^2
\end{equation}
parametrized by the constant $M$. This family includes the pure AdS spacetime for $M=0$. The solutions \eqref{static_sol} for $M \geq 1$ are known as BTZ black holes \cite{btz}. The solutions with $0<M<1$ have a conical singularity at $r=0$ which can be viewed as being due to the presence of a point particle in the origin, same as described in \cite{star}. Let us point out that the solution \eqref{static_sol} is a special case of the AdS-Schwarzschild family of solutions of the vacuum Einstein equations with the negative cosmological constant in $(d+1)$ dimensions
\begin{equation} \label{ads_schw}
ds^2=-\left(1-\frac{M}{r^{d-2}} + \frac{r^2}{\ell^2} \right)+\left(1-\frac{M}{r^{d-2}} + \frac{r^2}{\ell^2} \right)^{-1} dr^2 + r^2 d \omega^2_{S^{d-1}}.
\end{equation}
Note that in $d\geq 3$ the mass $M$ of the black hole can be arbitrarily small while in $d=2$ black holes exist only for $M\geq 1$, hence there is a mass gap between $\text{AdS}_{3}$ and the lightest black hole, which is an extreme BTZ black hole with $M=1$. The existence of this finite energy threshold for black hole formation in three-dimensional gravity will play a crucial role in our analysis.

 If we neglect the cosmological constant in \eqref{hamilton0} and \eqref{slicing} we get
\begin{eqnarray}
  \partial_r \tilde A  &=& -\kappa \,r \tilde A\, S, \label{eqA} \\
   \partial_r \delta &=& -\kappa\, r \,S \label{eqd} .
\end{eqnarray}
Multiplying \eqref{eqd} by $\tilde A$ and subtracting it with \eqref{eqA} one gets $\partial_r (\tilde A e^{-\delta})=0$, which together with the boundary conditions $\tilde A(\tau,0)=1$ and $\delta(\tau,0)=0$ leads to $\tilde A e^{-\delta} \equiv 1$. Therefore, Eq.\eqref{weq} is reduced to the radial wave equation in a flat spacetime. For any finite-energy solution of this equation, we can integrate Eq.\eqref{hamilton0} to get
\begin{equation*} %\label{A}
  \tilde A(\tau,r)=\exp\left(-\kappa \int_0^r S(\tau,r') r' dr'\right).
\end{equation*}
Thus, $\tilde A(\tau,r)$ can never reach zero and therefore no apparent horizon can develop during the evolution\footnote{Let us recall that the apparent horizon is defined by the condition $g^{\alpha\beta} \partial_{\alpha} r \partial_{\beta} r=0$, which for the metric \eqref{ansatz} gives $\tilde A=0$.}. This is a special case of a general result by Ida \cite{ida} who proved that
in three-dimensional gravity with zero cosmological constant the existence of trapped surfaces is excluded.

For further analysis, especially for numerical simulations, it is convenient to define dimensionless coordinates $(t,x)\in (-\infty,\infty)\times [0,\pi/2)$ by  $\tau=\ell t$ and $r=\ell \tan{x}$ and a new metric function $A=(1+r^2/\ell^2)^{-1} \tilde A$. Then, the metric \eqref{ansatz} takes the form
\begin{equation}\label{metric}
 ds^2=\frac{\ell^2}{\cos^2{\!x}} \left(-A e^{-2\delta} dt^2 + A^{-1} dx^2 + \sin^2{\!x}\, d\varphi^2\right)\,.
\end{equation}
Further, we use a unit of mass such that $\kappa=1$ and a unit of length such that $\ell=1$. We introduce auxiliary variables $\Phi$ and $\Pi$ defined by $\Phi=\phi'$ and $\Pi=A^{-1} e^{\delta} \dot \phi$ (using ${}'=\partial_x$ and ${}^\cdot=\partial_t$) and rewrite the wave equation \eqref{weq}  in the first order form
\begin{equation}\label{weq2}
  \dot \Phi = ( A e^{-\delta} \Pi)',\quad \dot \Pi = \frac{1}{\tan{x}} (\tan{x} A e^{-\delta} \Phi)'\,,
\end{equation}
 The system of Einstein equations (\ref{slicing}-\ref{momentum}) becomes
\begin{eqnarray}
  \delta' &=& -\sin x \cos x (\Pi^2+\Phi^2), \label{delta-x} \\
  A' &=& - \sin{x} \cos{x}\, A (\Pi^2+\Phi^2) +2 \tan{x}\, (1-A), \label{hamilton}\\
  \dot A &=& -2 \sin{x} \cos{x}\,  e^{-\delta} A^2 \Pi \Phi\,. \label{momentum-x}
\end{eqnarray}
To get a well-posed initial-value problem, the above equations need to be supplemented by suitable boundary conditions \cite{f2}. Smoothness at the center implies that near $x=0$ the fields behave as follows
   \begin{align}\label{bc0}
   \phi(t,x)&=f_0(t)+\mathcal{O}(x^2),\quad  \delta(t,x)=\mathcal{O}(x^2), \nonumber \\
   A(t,x) &=1+\mathcal{O}(x^2),
    \end{align}
where we used the normalization $\delta(t,0)=0$ so that $t$ be the proper time at the center. Near spatial infinity we assume that (using $\rho=\pi/2-x$)
 \begin{align}\label{bcinf}
   \phi(t,x)&=f_{\infty}(t) \rho^2+\mathcal{O}(\rho^4), \quad \delta(t,x)=\delta_{\infty}(t)+\mathcal{O}(\rho^4),\nonumber \\ A(t,x)&=1-M \rho^2 + \mathcal{O}(\rho^4),
 \end{align}
where the power series are uniquely determined by a constant $M$ and functions $f_{\infty}(t)$ and $\delta_{\infty}(t)$. These fall-off conditions ensure that the mass function  $m(t,x):=(1-A)/\cos^2{\!x}$ has a finite time-independent limit  $\lim_{x\rightarrow \pi/2} m(t,x)=M$ which can be interpreted as the total mass.

Our goal was to solve the initial-boundary value problem for the above equations for small initial data and determine an endstate of the evolution. Due to the presence of the mass gap, small enough perturbations of $\text{AdS}_{3}$ cannot evolve into a black hole, therefore we are left with dichotomy: global existence in time or a naked singularity formation. Before presenting the results of our investigations, let me describe our methodology which is a combination of analytic spectral methods and numerical simulations.

%%%%%%%%%%%%%%%%%%%%%%%%%%%%%%%%%%%%%%%%%%%%%%%%%%%%%%%%%%%%
%%%%%%% Analyticity strip method
%%%%%%%%%%%%%%%%%%%%%%%%%%%%%%%%%%%%%%%%%%%%%%%%%%%%%%%%%%%%
\section{The analyticity strip method}
The problem of global-in-time regularity of solutions is very hard for many nonlinear evolution equations. Some insight regarding a possibility of singularity formation can be gained from numerical simulations. Such simulations are notoriously difficult in the case of so called weakly turbulent dynamics which is characterised by the energy transfer from low to high frequencies. As the weakly turbulent solutions develop finer and finer spatial scales any numerical simulation eventually breaks down because of the loss of spatial resolution. Nevertheless, numerical simulations combined with a clever spectral technique called the analyticity strip method, introduced in \cite{ssf}, can help  predict if a singularity occurs in finite time or not. In this paragraph, we briefly present the main concepts behind this method.

The idea of the analyticity strip method is based on the relation between the analytic properties of a complex function and the large wavenumber behaviour of its Fourier transform. Consider a real function $u(t,x)$ to be a solution to an evolution equation starting from some real analytic initial data. Let $u(t,z)$ to be its extension to the complex z-plane. Typically, $u(t,z)$ will have  singularities in the complex plane. Let  $z_*=x_*+ i \rho$ be the location of the singularity which is closest to the real axis. Thus, the imaginary part $\rho$ measures  the width of a strip around the real axis  which is free of singularities. As the solution evolves, the width $\rho(t)$ evolves as well. If it goes to zero in a finite time, then the complex singularity hits the real axis and the solution $u(t,x)$ becomes singular, otherwise the solution remains real analytic. By monitoring the time dependence of $\rho(t)$ one can thus  predict or exclude a finite time blowup. What makes this method very useful in numerical calculations, is the fact that the value of $\rho$ is encoded in the behaviour of Fourier coefficients for large wavenumbers
\begin{equation}\label{fourier}
\hat{u_k} \sim |k|^{\mu} e^{- \rho k} e^{i x_* k} \text{ for } k \to +\infty,
\end{equation}
where $\mu$ is connected with the order of the pole at $z_*$ and $k$ is the wavenumber. The derivation of \eqref{fourier}  can be found in \cite{ckp}. It is assumed that singularities are isolated points. It follows from \eqref{fourier} that $\rho (t)$ can be determined  by fitting an exponential decay to the tail of the numerically computed Fourier spectrum.

As every numerical method, the analyticity strip method  has some limitations. When the width of analyticity becomes too small in comparison with the mesh size, then the uncertainty of the position of the singularity is of the same order as its distance from the real axis and the results are no longer reliable. Such a situation occurs when $\rho(t)$ decreases to zero in time. When it stays bounded away from zero, then the time integration can be carried on for arbitrarily long times.  Another problem is due to the fact that numerically we can include only finitely many wavenumbers up to some cut-off $k_{max}$. Fortunately, it turns out that the fit of $\rho$ is only weakly dependent on the ultraviolet cut-off, provided that $k_{max}$ is large enough.

Let us illustrate the analyticity strip method with an example. This example is academic because $\rho(t)$ can be calculated analytically but it helpful in understanding different notions of stability.
\vskip 0.2cm
\noindent \textbf{Example} Consider the following hyperbolic equation for $u(t,x)$
\begin{equation*}
\partial_t u= x \partial_x u + \alpha u^2,
\end{equation*}
where $\alpha$ is a nonegative constant. Let us take an initial datum
\begin{equation*}
u(0,x)= \frac{\epsilon}{1+x^2},
\end{equation*}
which has complex singularities at $x=\pm i$. The corresponding solution can be easily obtained by  the method of characteristics
\begin{equation} \label{ex_sol}
u(t,x)=\frac{\epsilon}{1+e^{2t} x^2-\alpha \epsilon t}.
\end{equation}
Its Fourier transform in space is given by
\begin{equation}\label{sol-fourier}
\hat u (t,k)=\frac{\epsilon \pi e^{-t}}{\sqrt{1-\epsilon \alpha t}} H(k) \exp (-k \underbrace{e^{-t} \sqrt{1-\epsilon \alpha t}}_{\rho(t)})+ (k \leftrightarrow -k).
\end{equation}
 It follows from \eqref{ex_sol} that for $\alpha>0$ the solution blows up at $x=0$  at time $T=1/\epsilon \alpha$. In \eqref{sol-fourier} this is reflected in $\rho(T)=0$.
 For $\alpha=0$ the solution is globally regular and $\rho(t)=e^{-t}$.  This exponential decay of the width of analyticity is characteristic for the weakly turbulent dynamics and has important consequences regarding the stability. To see this, consider the Sobolev norms  given by
$$
||u(t)||^2_{\dot H_s}=\int \limits_{-\infty} ^{+\infty} (\partial_x^s u)^2 dx = e^{(2s-1)t} \, ||u(0)||^2_{\dot H_s}\,.
$$
 For $s=0$ the norm goes to zero as $t\rightarrow \infty$ which means that the zero solution is $L^2$ asymptotically stable. However, for $ s>\frac{1}{2}$ the norms grow exponentially in time which means instability. This example illustrates that stability properties may be sensitive to the choice of the norm which measures the size of perturbations.

%%%%%%%%%%%%%%%%%%%%%%%%%%%%%%%%%%%%%%%%%%%%%%%%%%%%%%%%%%%%
%%%%%%% Spectral properties
%%%%%%%%%%%%%%%%%%%%%%%%%%%%%%%%%%%%%%%%%%%%%%%%%%%%%%%%%%%%

\section{Spectral properties}

The first step in determining stability of an equilibrium solution of an evolution equation is to analyze its linear stability. It is done by linearizing the equations around this solution. In the case at hand, we need to linearize the system \eqref{weq2}-\eqref{momentum-x} around the AdS$_3$ solution $A=1$, $\delta=0$, $\phi=0$. This gives
just the wave equation on the pure $\text{AdS}_{3}$ background
\begin{equation}\label{lin-pert}
\ddot \phi +L \phi=0\,,\qquad L=-\tan^{-1}{x}\, \partial_x (\tan{x} \, \partial_x)\,.
\end{equation}
This equation is a special case of the master equation for various linear perturbations of AdS: scalar, electromagnetic and gravitational, which were studied in detail by Ishibashi and Wald \cite{wi}.
The operator $L$
  is self-adjoint on the Hilbert space $L^2([0,\pi/2], \tan{x}\, dx)$. Eq.\eqref{lin-pert} can be solved by the standard separation of variables. One finds that the eigenvalues of $L$ are given by $\omega_k^2=(2+2k)^2$, where $k=0,1,\dots$, and the corresponding eigenfunctions are the Jacobi polynomials in $\cos{2x}$
\begin{equation*} %\label{eigenF}
e_k(x)=2 \sqrt{k+1} \, \cos^2{x}\, P_k^{0,1}(\cos{2x}).
\end{equation*}
These eigenfunctions are orthonormal with respect to the inner product
$$
(f,g):= \int \limits_0^{\pi/2} f(x) g(x) \tan{x}\, dx.
$$
The distance between the successive eigenfrequencies  is constant which means that, in the PDE terminology, the spectrum if fully resonant. It is believed that this property lies at the root of the weakly turbulent instability of AdS \cite{br,bll,dhs}. In order to quantify the shift of energy from low to high frequencies we introduce the projections
\begin{equation} \label{projections}
\Phi_k := (\sqrt{A} \Phi,e'_k), \quad \Pi_k:=(\sqrt{A}\Pi,e_k).
\end{equation}
Then, using the fact that the derivatives of eigenfunctions fulfill the orthogonality relation $(e'_i,e'_j)=\omega_i^2 \delta_{ij}$, the total mass can be expressed as the Parseval sum:
\begin{equation}\label{parseval}
   M=\frac{1}{2} \int \limits_0 ^{\frac{\pi}{2}} \left( A \Phi^2 + A \Pi^2 \right) \tan x dx=\sum_{k=0}^{\infty} E_k=\sum_{k=0}^{\infty} \Pi_k ^2+\omega_k ^{-2} \Phi_k^2,
\end{equation}
where $E_k$ can be interpreted as the energy concentrated in the k-th mode.

%%%%%%%%%%%%%%%%%%%%%%%%%%%%%%%%%%%%%%%%%%%%%%%%%%%%%%%%%%%%
%%%%%%% Numerical methods
%%%%%%%%%%%%%%%%%%%%%%%%%%%%%%%%%%%%%%%%%%%%%%%%%%%%%%%%%%%%

\section{Description of numerical methods}
In this section I describe the numerical methods used in \cite{bj} to solve the system \eqref{weq2}-\eqref{momentum-x}.

 The system \eqref{weq2}-\eqref{momentum-x} is overdetermined, so we have a choice of updating the metric function $A(t,x)$ either by the momentum constraint \eqref{momentum-x} or the Hamiltonian constraint \eqref{hamilton}. These two choices lead to so called free or constrained evolution schemes, respectively. We used the constrained evolution scheme because it is more stable.
 We use the standard method of lines which in our case works as follows: we approximate spatial derivatives in equations \eqref{weq2} using a fourth-order finite difference scheme and then we use the classical fourth-order Runge-Kutta method to integrate equations in time. After every elementary time step we update metric variables $A$ and $\delta$ by solving equations \eqref{delta-x} and \eqref{hamilton}. Additionally, we impose the boundary conditions \eqref{bc0} and \eqref{bcinf}. To remove high frequency instabilities we add Kreiss-Oliger dissipation. The momentum constraint \eqref{momentum-x} was used to monitor the accuracy of  computations. We also performed convergence tests , which for finite difference methods work as follows.  Given  a numerical solution $\Phi_n$ computed on a grid with $2^n$ grid points, one defines the convergence factor
\begin{equation}
Q_n = \frac{||\Phi_n-\Phi_{n+1}||}{||\Phi_{n+1}-\Phi_{n+2}||}
\end{equation}
where $||.||$ is any suitable norm, for instance the $\ell_2$ norm as used here. By the Richardson lemma, for a convergent finite-difference method of order $N$ this quantity is equal to $2^N$, hence in our case it should be $2^4$.
 \begin{figure} [h]
 \centering
 \includegraphics[width=0.5\textwidth,angle=270]{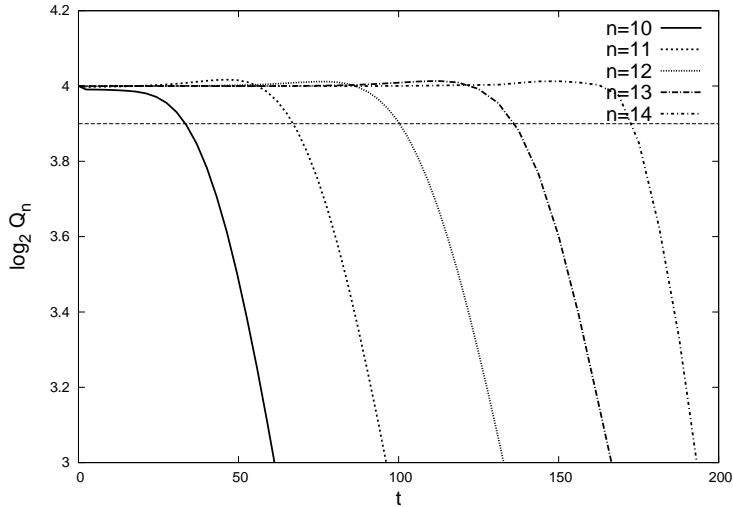}
  \caption{Runs for convergence tests were performed on grids with $2^n$ points for $n=10, \ldots, 16$ (for the initial data   \eqref{data} with $\varepsilon=0.3$). The plot shows $\log_2 Q_n$ as a function of time.  The horizontal dashed line depicts the deviation by $7\%$ from the expected value $4$. It is seen that the time after which the convergence is lost scales approximately  linearly with $n$.}
 \label{conv}
\end{figure}
All numerical results presented in this work were obtained for the initial data of the form
  \begin{equation}\label{data}
  \phi(0,x)=\varepsilon \exp(-\tan^2{\!x}/\sigma^2)\,,\quad \dot\phi(0,x)=0
  \end{equation}
   with width $\sigma=1/32$ and numerous small amplitudes $\varepsilon$. The results of convergence test for these data are presented in Fig. \ref{conv}.

Numerical simulations of turbulent phenomena are very difficult due to the fact that the energy concentrates on smaller and smaller spatial scales and after some time solutions cannot be  properly resolved. This  causes the loss of convergence after some 'reliability time', as shown  in Fig. \ref{conv}. On the one side, we need fine spatial resolution to resolve small scales developing during evolution; on the other side, the computational complexity of the numerical method is $\mathcal{O} (k^2)$, where $k$ is the number of grid points which means that doubling the number of grid points makes the simulation time roughly four times longer. To achieve a balance between the  spatial resolution and  duration of simulations, we monitor results obtained on grids with $2^n$ and $2^{n-1}$ points and refine the entire spatial grid when the difference exceeds some tolerance level. This method is numerically expensive but very stable and provides an error control. In our simulations we started from grids with $2^{12}$ points and allowed for four levels of refinement. By extrapolation, we estimate that the reliability time for the smallest amplitude $\varepsilon=0.3$ and the highest resolution of $2^{16}$ points is $230$. Consequently, in our analysis we did not use the data simulated for longer times.

%%%%%%%%%%%%%%%%%%%%%%%%%%%%%%%%%%%%%%%%%%%%%%%%%%%%%%%%%%%%
%%%%%%% Numerical results
%%%%%%%%%%%%%%%%%%%%%%%%%%%%%%%%%%%%%%%%%%%%%%%%%%%%%%%%%%%%

\section{Numerical results for small perturbations of AdS$_3$}
 In \cite{br, jrb} it was found that small perturbations of AdS$_{d+1}$ for $d\geq 3$  generically evolve into a black hole. This instability of AdS has been conjectured to be caused by the resonant transfer of energy to high frequencies.
 As I explained above, in three dimensions the black hole formation is excluded for small enough perturbations, hence we face a dichotomy: global-in-time regularity or singularity formation in finite time. Here I summarize the results of numerical simulations \cite{bj} which indicate that solutions remain globally regular.

  We numerically solved the initial-boundary value problem for the system \eqref{weq2}-\eqref{hamilton} with the boundary conditions \eqref{bc0},\eqref{bcinf} and the initial data  \eqref{data}. The amplitudes $\varepsilon$ were chosen so that the total mass $M\ll 1$.

\begin{figure}[h]
\centering
 \includegraphics[width=0.5\textwidth,angle=270]{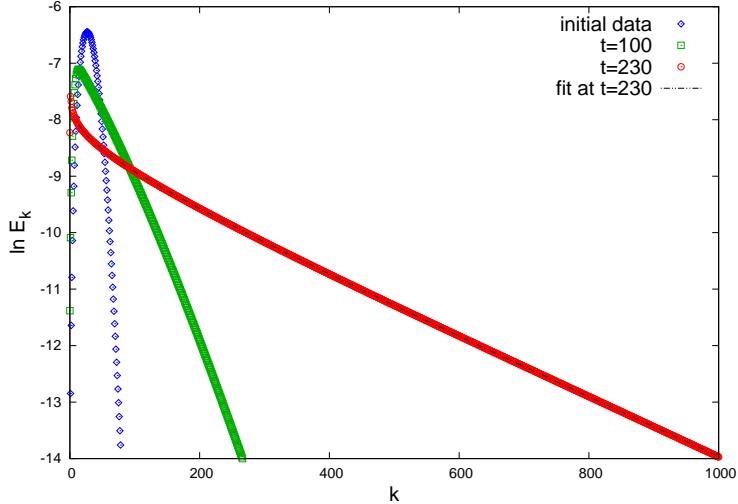}
  \caption{Energy spectra at $t=0$, $t=100$ and $t=230$ for $\varepsilon=0.3$. The total mass of this data is equal to $M=0.044$. We present the data in lin-log scale to better demonstrate the exponential decay of the tail of the spectrum for high frequencies. The fit of the formula (\ref{fit}) in the interval $10<k<1000$ to the data at $t=230$  is depicted by the black dotted line.}
 \label{spectrum}
\end{figure}

The most important quantitative characterisctics of any turbulent phenomenon is the energy spectrum. To get the distribution of the total energy between modes of a linearized equation, we have to project $\Phi$ and $\Pi$ using \eqref{projections} at some chosen moment of time $t$ and from $\eqref{parseval}$ calculate $E_k$ for $k < k_{max}$. In Fig. \ref{spectrum} we present results of this procedure for $\varepsilon=0.3$. It is seen that  higher and higher frequencies are excited during evolution (comparing this with an analogous plot of the energy spectrum in \cite{prague}, we see that the energy transfer is  slower in higher dimensions). Approximately  $40$ modes are excited initially, whereas at $t=230$ all modes up to $k_{max}=1000$ participate in the evolution.

\begin{figure}[h]
 \centering
 \includegraphics[width=0.5\textwidth,angle=270]{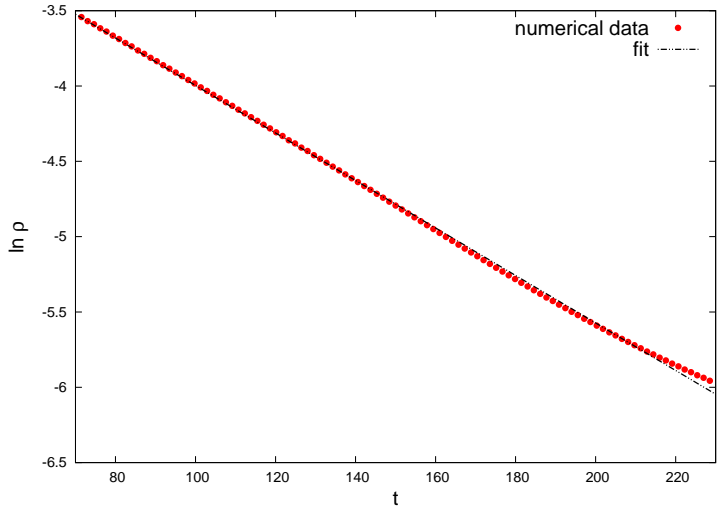}
  \caption{Fitting the formula \eqref{fit} to the energy spectra gives the time dependence of the width of analyticity $\rho(t)$, which is shown in this plot. Simulations were performed for the same initial data as in Fig.\ref{spectrum}. The fit of the exponential decay, given by \eqref{delta}, (depicted by the dashed line) gives $\rho_0=0.09, T=63.4$.  Changing the fitting interval for $k$ when fitting \eqref{spectrum} to the energy distribution, does not influence the results for $\rho_0$ and $T$ much.}
 \label{log_dec}
\end{figure}

According to the analyticity strip method, energy for large wave numbers behaves as follows
 \begin{equation}\label{fit}
   E_k(t)=C(t)\,k^{-\beta(t)} e^{-2 \rho(t) k},
 \end{equation}
where we assume that parameters $C,\beta,\rho$ are time-dependent. Fig. \ref{spectrum} confirms that the energy spectrum exponentially decays for large $k$. By fitting the formula \eqref{fit} to the data, we determined the time dependence of parameters. We focus on the parameter $\rho$ because it serve as the key indicator of the loss of regularity. We find that $\delta(t)$   is always bounded away from zero and after some time it decreases  exponentially with a characteristic decay time $T$
 \begin{equation}\label{delta}
 \rho(t)=\rho_0 \, e^{-t/T}\,.
  \end{equation}
 The evidence for \eqref{delta} is shown in Fig. \ref{log_dec}. The constant $\rho_0$ is independent of the amplitude of  initial data and the time $T$ is proportional to $\varepsilon^{-2}$. The width of analyticity is of the order of hundred grid steps at the reliability time, which reassures us that the results are trustworthy. In Fig. \ref{log_dec} we present also a fit to the numerical data. A good agreement suggests that this behaviour could be extrapolated to longer times beyond $230$, however higher resolution simulations would be helpful to feel more confident about this extrapolation. Anyway, we get no indication that the solution becomes singular.
The exponential decay of the width of analyticity strip indicates the gradual loss of regularity of solutions.

\begin{figure} [h]
\centering
 \includegraphics[width=0.5\textwidth,angle=270]{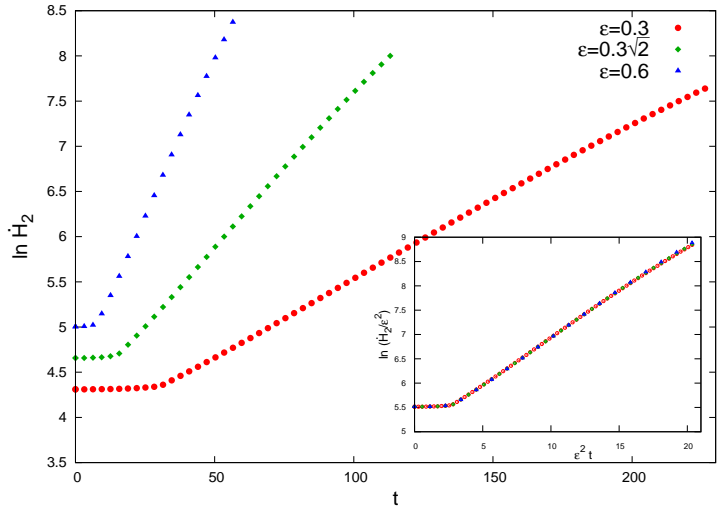}
  \caption{This fgure presents time evolution of the $L^2$-norm of the second spatial derivative $\dot H_2=||\phi''(t,x)||_2 $ for the initial data with three different amplitudes: $\varepsilon=0.3, 0.3\sqrt{2}, 0.6$. $\dot H_2$ rapidly oscillates in time, so for clarity, only local maxima in time are plotted. Results are presented in lin-log scale to show the exponential growth of the norm. In the inset,  the curves corresponding to three different amplitudes $\varepsilon$ are rescaled according to $t \rightarrow \varepsilon^2 t$ and $\dot H_2 \rightarrow \dot H_2/\varepsilon^2$. After such rescaling, all three curves coincide. Despite of the growth of the Sobolev norm, we do not get any sign that the solution loses smoothness.}
 \label{norm}
\end{figure}

We also studied the behavior of Sobolev norms to test the stability of the solutions.
Fig. \ref{norm} gives  numerical evidence for instability of $\text{AdS}_{3}$ where we study the behaviour of the maxima of the second homogenous Sobolev norm $\dot H_2$ defined as $\dot H_2=||\phi''(t,x)||_2 $. It turns out that after an initial flat period, the maxima of $\dot H_2$ start growing exponentially, as could be guessed from \eqref{fit} by dimensional analysis. The time of onset of the exponential growth scales as $\varepsilon^{-2}$ as shown on the an inset in Fig. \ref{norm}. Higher Sobolev norms exhibit a similar exponential growth.

%%%%%%%%%%%%%%%%%%%%%%%%%%%%%%%%%%%%%%%%%%%%%%%%%%%%%%%%%%%%
%%%%%%% Podsumowanie
%%%%%%%%%%%%%%%%%%%%%%%%%%%%%%%%%%%%%%%%%%%%%%%%%%%%%%%%%%%%

\section{Final remarks}

Numerical simulations for initially small scalar perturbations of $\text{AdS}_{3}$ performed in \cite{bj} indicate that  such perturbations give rise to solutions which remain  globally regular in time. However, their higher Sobolev norms grow exponentially, which means that the solutions gradually lose smoothness by developing finer and finer spatial scales as $t \rightarrow \infty$. Such process is called  weak turbulence and has been known in fluid dynamics (for example, the proof of weak turbulence for the  incompressible Euler equations in two spatial dimensions can be found in \cite{yu, bbz}). It is widely believed that weak turbulence is a common feature for  many nonlinear wave equations in bounded domains, however the rigorous results in this matter are extremely rare \cite{ckstt, gg,cf,gt}. For Einstein equations the weakly turbulent dynamics is possible only in three dimensions, because in higher dimensions the energy transfer is cut off by a black hole formation.

We discussed only small perturbations of $\text{AdS}_{3}$ with the total mass $M \ll 1$ but we observed very similar behaviour for a  variety of subcritical initial data. We believe that all solutions with $M < 1$ are globally regular in time but higher resolution numerical simulations would be very helpful in validating this conjecture.

Another interesting open problem to be adressed is the threshold for a black hole formation at $M=1$.  Such a finite energy threshold for a blowup is typical for nonlinear wave equations in critical dimensions. Some light on the near critical dynamics was shed in numerical studies done in \cite{pc}. Nevertheless, the critical solution remains still not well understood \cite{carsten}.
% Furthermore, it is not obvious that every perturbation of $\text{AdS}_{2+1}$ with $M >1$ will evolve into a black hole. This region is also worth investigating.

 % Acknowledgemnts --------------------------------------------------------------------------------------------------------------------------------------
 %%%%%%%%%%%%%%%%%%%%%%%%%%%%%%%%%%%%%%%%%%%%%%%%%%%%%%%%%%%%
\vskip 0.1cm \noindent \emph{Acknowledgments:}
It is a pleasure to thank Piotr Bizo\'n, my advisor and collaborator, for teaching me many of the things described here and for critical comments on the draft. This work was supported by the NCN grant DEC-2012/06/A/ST2/00397 and  Foundation for Polish Science under the MPD Programme ÔÔGeometry and Topology in Physical Models.ÕÕ The computations were performed at the Academic Computer Centre Cyfronet AGH using the PL-Grid infrastructure.

\end{document}